\documentclass[aps,prl,twocolumn,showpacs,letterpaper]{revtex4-1}
\pdfoutput=1
\usepackage{times}
\usepackage{epsfig}
\usepackage{amsmath}
\usepackage{amssymb}
\usepackage{wasysym}
\usepackage[colorlinks=true,linkcolor=blue, citecolor=red]{hyperref}

\newcommand{\bs}{\boldsymbol}

\begin{document}

\title{Topological Order and Semions in a Strongly Correlated Quantum Spin Hall Insulator}
\author{Andreas R\"uegg and Gregory A. Fiete}
\affiliation{Department of Physics, The University of Texas at Austin, Austin, Texas 78712, USA}

\date{\today}

\begin{abstract}
We provide a self-consistent mean-field framework to study the effect of strong interactions in a quantum spin Hall insulator on the honeycomb lattice. We identify an exotic phase for large spin-orbit coupling and intermediate Hubbard interaction. This phase is gapped and does not break any symmetry. Instead, we find a four-fold topological degeneracy of the ground state on the torus and fractionalized excitations with semionic mutual braiding statistics. Moreover, we argue that it has gapless edge modes protected by time-reversal symmetry but a trivial $Z_2$ topological invariant. Finally, we discuss the experimental signatures of this exotic phase. Our work highlights the important theme that interesting phases arise in the regime of strong spin-orbit coupling and interactions.
\end{abstract}



\maketitle

{\it Introduction --}
Time-reversal invariant topological insulators (TIs) \cite{Hasan:2010} are known for their robust and peculiar response to topological defects. For example, certain lattice dislocations in a weak TI support gapless one-dimensional helical modes \cite{Ran:2009}. Threading a TI with a $\pi$-flux tube (in units of $\hbar c/e$) leads to spin-charge separated excitations in the two-dimensional quantum spin Hall (QSH) insulator \cite{Lee:2007,Qi:2008,Ran:2008} and the ``wormhole effect" in a three dimensional strong TI \cite{Rosenberg:2010b}. In TI hybrid structures, even more exotic behavior is expected: for example, a vortex in an s-wave superconductor deposited on a strong TI binds a single Majorana mode to its core \cite{Fu:2008}. 

The above-mentioned intriguing properties result from the topologically non-trivial electronic structure obtained in the presence of spin-orbit coupling but with the electron-electron interactions treated on a single-particle level. Recently, there has been an increasing effort to analyze the regime where the single-particle picture (partly) breaks down \cite{Shitade:2009,Dzero:2010,Chaloupka:2010,Wang:2011,Pesin:2010,Kargarian:2011}. The study of these ``correlated TIs'', such as certain heavy-electron systems \cite{Dzero:2010} or 5d-based transition metal oxides \cite{Shitade:2009,Pesin:2010,Kargarian:2011}, raises the question how the underlying topologically non-trivial band structure affects the fate of the physics of the interacting system, e.g., in magnetic insulators \cite{Wang:2011} or spin liquids \cite{Pesin:2010}.

In this article, we elaborate on a rather general but unique aspect of correlated TIs: the possibility of novel and exotic excitations in the interacting limit which have their antecedent in the characteristic properties of the non-interacting system. Indeed, emergent quasiparticles in a correlated TI which are associated with topological defects of an order parameter \cite{Grover:2008} or an emergent gauge field \cite{Ran:2008} have been proposed previously. 
In particular, it has been shown \cite{Ran:2008} that a QSH insulator coupled to a dynamical $Z_2$ gauge field (denoted by QSH$^*$ in the following) supports bosonic excitations which carry a fraction of the electronic quantum numbers and have semionic mutual braiding statistics. Yet, a scheme of how such a fractionalized phase can emerge from a microscopic interacting Hamiltonian has been missing. In this article, we provide a physical lattice model where the QSH$^*$ insulator is found within a self-consistent mean-field analysis. Our approach reveals especially rich physics in the regime where both the spin-orbit coupling and the interactions are strong. Notably, we identify a topological degeneracy of the ground state and demonstrate the existence of fractionalized excitations. 

\begin{figure}
\centering
\includegraphics[width=1\linewidth]{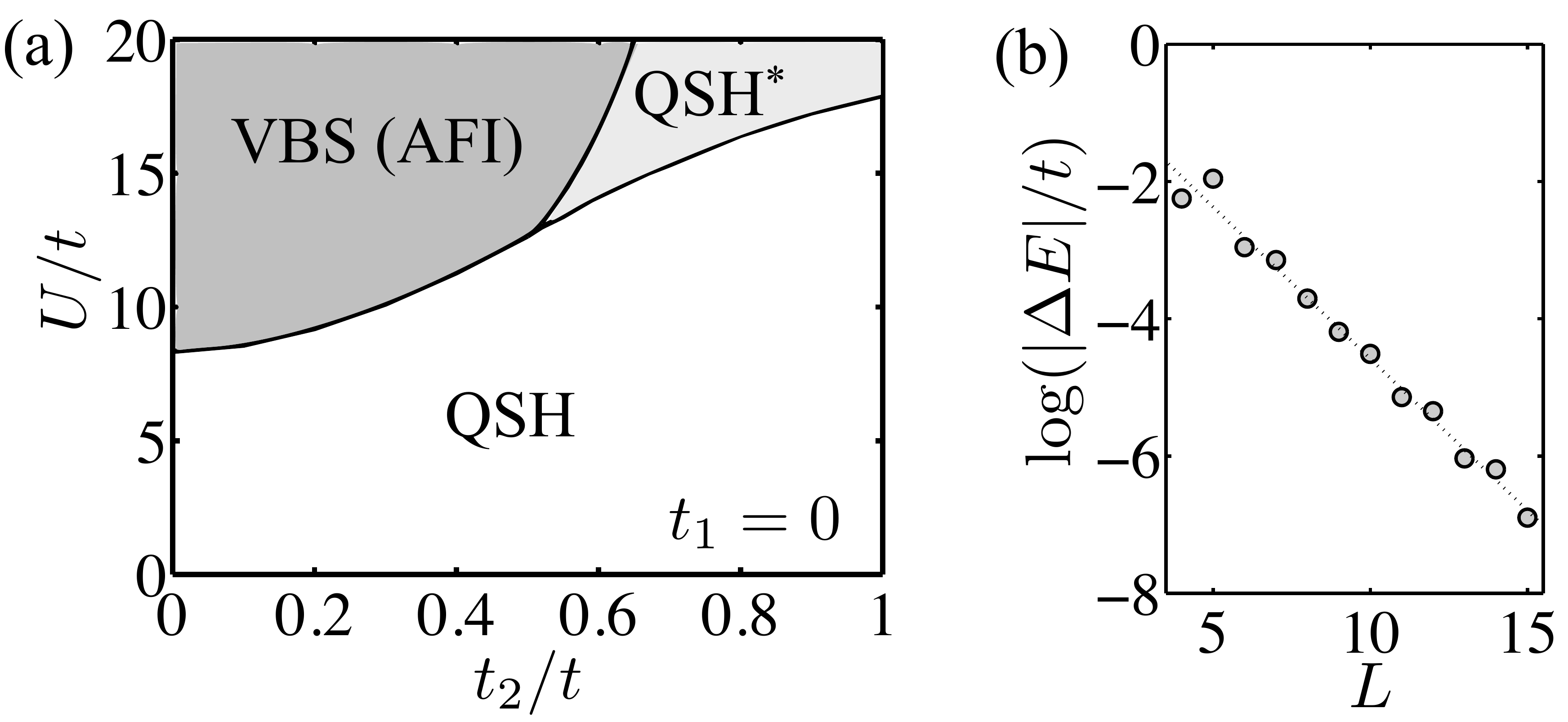}
\caption{(a) Phase diagram obtained in the $Z_2$-mean-field approximation. QSH is a quantum-spin Hall insulator and VBS a valence bond solid. The focus of the current article is the strongly correlated QSH$^*$ at intermediate $U$ and large spin-orbit coupling $t_2$. (b) Topological ground state degeneracy in the QSH$^*$: Logarithm of the energy difference between solutions with and without a global $Z_2$ flux on a $L\times L$ torus for $U=26t$ and $t_2=t$.}
\label{fig:PD_TO}
\end{figure}

{\it Model --}
We consider the half-filled Hubbard model on the honeycomb lattice proposed for a single-layer of the layered compound Na$_2$IrO$_3$ in Ref.~\cite{Shitade:2009}:
\begin{equation}
H=\sum_{i,j}\sum_{\alpha,\beta}t_{ij}^{\alpha\beta}c_{i\alpha}^{\dag}c_{j\beta}^{}+U\sum_in_{i\uparrow}n_{i\downarrow}.
\label{eq:model}
\end{equation}
Here, $c_{i\alpha}^{(\dag)}$ (creates) annihilates an electron in a spin-orbital coupled pseudo-spin 1/2 state $\alpha$ (hereafter called spin) which denotes the low-energy doublet of the spin-orbit coupled $t_{2g}$ orbitals at the Ir atom $i$. The nearest-neighbor hopping $t_{i,j}^{\alpha,\beta}=-\sigma^{0}_{\alpha\beta}t$ is spin-independent while the second-neighbor hopping is complex and a function of the spin: $t_{i,j}^{\alpha,\beta}=-t_1\sigma_{\alpha\beta}^{0}+it_2\sigma^{w}_{\alpha\beta}$. Here, $w=x,y,z$ depends on the direction from $i$ to $j$ \cite{Shitade:2009}, $\sigma^0$ is the identity and $\sigma^{x,y,z}$ are the Pauli matrices. $t_2$ characterizes the atomic spin-orbit coupling which, as opposed to the $S_z$-conserving model \cite{Rachel:2010, Soriano:2010, He:2011, Zheng:2010, Hohenadler:2011, Lee:2011, Imada:2011}, leads to a full breakdown of the spin-rotation symmetry and some additional degree of magnetic frustration \cite{Chaloupka:2010}. The non-interacting model is in the symplectic symmetry class (AII) \cite{Schnyder:2008} and realizes the QSH insulator at half-filling.

{\it Phase diagram:} 
Figure \ref{fig:PD_TO}(a) shows the variational phase diagram of the model Eq.~\eqref{eq:model} obtained in the mean-field approximation of the recently introduced $Z_2$-slave-spin theory \cite{Huber:2009, Ruegg:2010b, Schiro:2010, Schiro:2011}, see below. For small $U$ the QSH state is stable. Because our variational approach preserves the time-reversal symmetry, we can not capture the expected magnetically ordered states in the large $U$ limit. Instead, we find a valence-bond solid (VBS) with increasing $U$ at small $t_2$. The VBS can be considered as the closest relative to an antiferromagnetic insulator (AFI) with spontaneously broken time-reversal symmetry \cite{Shitade:2009}. Indeed, on the honeycomb lattice, VBS states are close in energy to the AFI \cite{Reuther:2011, Albuquerque:2011}. The focus of this study is the intermediate phase at large spin-orbit coupling $t_2$ where we find the strongly correlated QSH$^*$ insulator. (The spin-liquid phase recently found in a small window around vanishing spin-orbit coupling and intermediate interactions \cite{Meng:2010,Hohenadler:2011,Zheng:2010,Wu:2011} is not captured by our approach.) As discussed below, the QSH$^*$ phase survives beyond the uniform mean-field treatment.

{\it $Z_2$ representation --} 
The starting point of our formal discussion of the QSH$^*$ is a representation of the model Eq.~\eqref{eq:model} in an enlarged Hilbert space which consists of fermionic pseudoparticles denoted by $f_{i\alpha}$ and auxiliary $S=1/2$ slave spins $s_i$ \cite{Huber:2009}. This representation makes use of the particle-hole symmetry of the Hubbard interaction at half-filling and the Hamiltonian takes the form
\begin{equation}
{\mathcal H}=\frac{1}{S^2}\sum_{i,j}\sum_{\alpha,\beta}t_{ij}^{\alpha\beta}s_i^xs_j^xf_{i\alpha}^{\dag}f_{j\beta}^{}+\frac{U}{4S}\sum_i\left(S+s_i^z\right).
\label{eq:modelSS}
\end{equation}
Equation~\eqref{eq:modelSS} is a faithful representation of the original model Eq.~\eqref{eq:model} in the subspace where all the local operators $u_i:=(-1)^{s_i^z-S+n_i}$ (with $n_i=\sum_{\alpha}f_{i\alpha}^{\dag}f_{i\alpha}^{}$) act as identity, $u_i=1$. Note that $u_i$ performs a local $Z_2$ gauge transformation: $u_if_{i\alpha}^{(\dag)}u_i=-f_{i\alpha}^{(\dag)}$ and $u_is_i^xu_i=- s_i^x$. Hence, the physical subspace is the gauge-invariant subspace. Given a state $|\Psi\rangle$ in the enlarged Hilbert space, a physical state can be obtained by projection as long as $s=\langle\Psi|\mathcal{P}^2|\Psi\rangle\neq0$: $|\Phi\rangle=s^{-1/2}\mathcal{P}|\Psi\rangle$ where $\mathcal{P}=\prod_i(1+u_i)/2$. Note that the pseudoparticles $f_{i\alpha}$ carry both spin and charge of the original electron and no {\it ad hoc} assumption of their separation has been made. Nevertheless, we will see that spin-charge separated quasiparticles emerge in the strongly correlated limit.

{\it Mean-field theory --}
To proceed we study the model Eq.~\eqref{eq:modelSS} in a mean-field approximation: assuming a product form in pseudoparticles and slave spins we find a non-interacting problem for the fermions, 
$
H_f=\sum_{i,j,\alpha,\beta}t_{ij}^{\alpha\beta}g_{ij}f_{i\alpha}^{\dag}f_{j\beta}^{}
$
and the transverse-field Ising model for the slave spins, 
$
H_s=-1/S^2\sum_{(i,j)}J_{ij}s_i^xs_j^x+U/(4S)\sum_is_i^z,
$
supplemented with the self-consistency equations $g_{ij}=\langle s_i^xs_j^x\rangle/S^2$ and $J_{ij}=-\sum_{\alpha\beta}\langle t_{ij}^{\alpha\beta}f_{i\alpha}^{\dag}f_{j\beta}^{}+{\rm h.c.}\rangle$. The mean fields $g_{ij}$ and $J_{ij}$ are real and change sign under a gauge transformation: for $f_{i\sigma}^{(\dag)}\rightarrow\epsilon_i f_{i\sigma}^{(\dag)}$, $s_i^{x}\rightarrow\epsilon_i s_i^x$ with $\epsilon_i=\pm1$, the mean-fields transform according to $g_{ij}\rightarrow\epsilon_ig_{ij}\epsilon_j$ and $J_{ij}\rightarrow\epsilon_iJ_{ij}\epsilon_j$. Mean-field solutions which are related by a gauge transformation describe the same physical state after projection. 

By comparing the energies for different solutions of the self-consistency equations we find the bulk phase diagram shown in Fig.~\ref{fig:PD_TO}(a). To compute $g_{ij}$ we have used a 4-site cluster-mean field approximation which preserves the variational character of the total energy. The VBS is specified by $g_{ij}\neq0$ on one of the three nearest-neighbor bonds and zero otherwise. Both QSH and QSH$^*$ phases do not break any symmetry and we can choose $g_{ij}, J_{ij}>0$. They are distinguished by the fact that the ground state of the self-consistent Ising model is either in the ferromagnetic phase with $Z=\langle s_i^x\rangle^2/S^2>0$ for QSH or in the paramagnetic phase with $Z=0$ for QSH$^*$.

To test the stability of the uniform mean-field solution we have investigated inhomogeneous solutions on tori with $L_1$ unit cells in the ${\bs a}_1$ direction and $L_2$ unit cells in the ${\bs a}_2$ direction. The transverse field Ising model has been solved in a semiclassical (large $S$) approximation which is easily generalized to inhomogeneous configurations $\{J_{ij}\}$ (but it violates the variational character). The central observation is that in the QSH$^*$ there are self-consistent solutions where closed paths $\mathcal{C}$ encircling a $Z_2$ flux exist: 
$
\prod_{(ij)\in\mathcal{C}}{\rm sign}(g_{ij})=-1.
$ 
(On the other hand, in the QSH phase at small $U$, such solutions do not exist because $g_{ij}\approx \langle s_i^x\rangle\langle s_j^x\rangle/S^2$. This requires that along any loop an even number of bonds are negative.)

\begin{figure}
\centering
\includegraphics[width=1\linewidth]{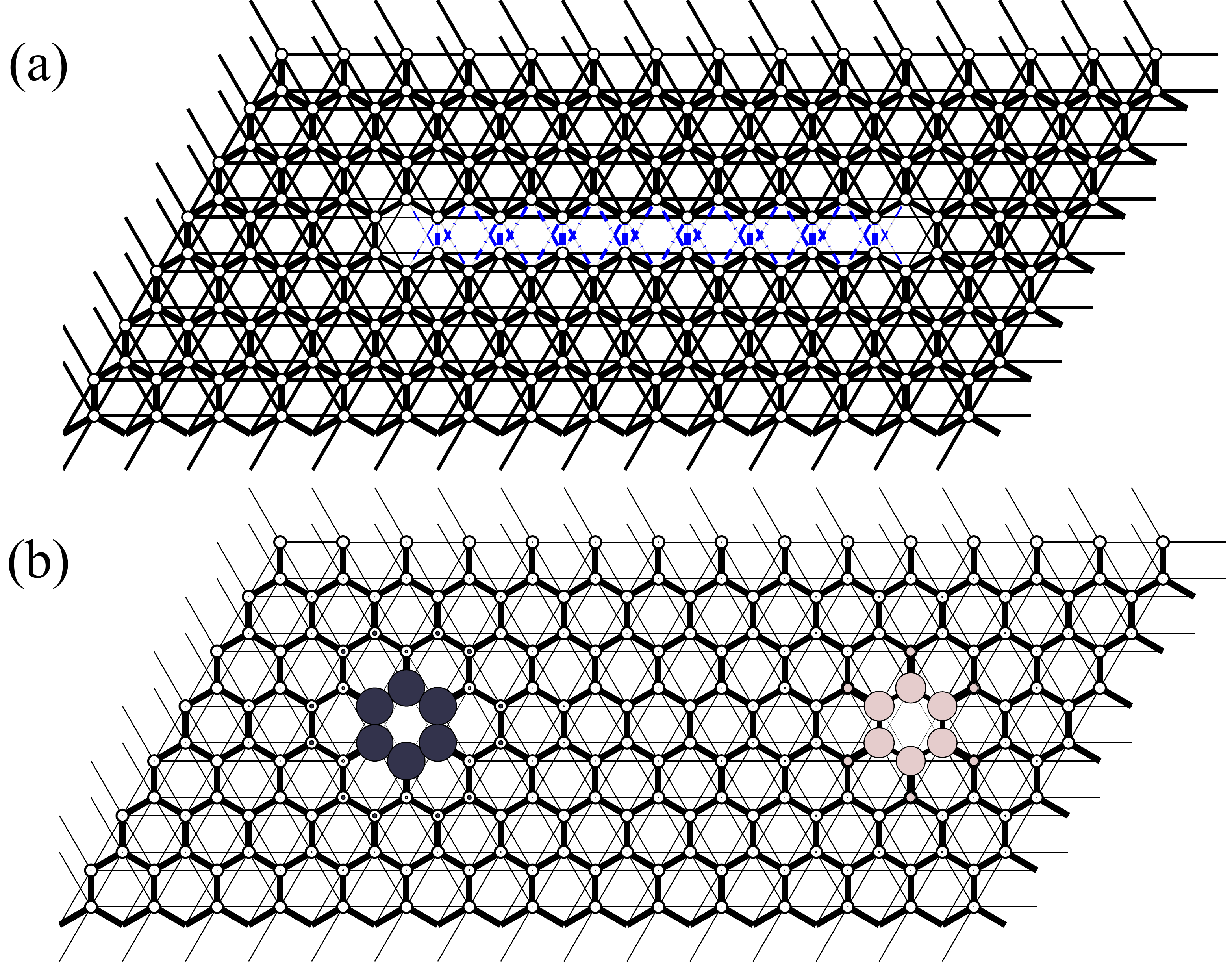}
\caption{(Color online.) Invisibility of the string in the QSH$^*$. (a) The (unphysical) bond variables $g_{ij}$ in the presence of two $Z_2$ flux excitations (doublon-holon pair). Dash-dotted (blue) lines are negative. (b) The kinetic bond energy $g_{ij}J_{ij}$ for the same configuration as in (a) along with the deviation of the charge density from half-filling (dark disks denote excess and bright deficit charge). The string is invisible in physical quantities. The parameters are $U=21t$, $t_1=0.2t$ and $t_2=0.7t$.}
\label{fig:fluxpair}
\end{figure}

There are profound consequences of the above observation. In particular, we find that the ground state of the QSH$^*$ is four-fold degenerated in the thermodynamic limit as $L_{1,2}\rightarrow\infty$, see Fig.~\ref{fig:PD_TO}(b). The four different ground states are characterized by the presence or absence of two global $Z_2$ fluxes. This degeneracy is robust against local perturbations of the system and is therefore a {\it topological degeneracy}. Numerically, the bond enegies $g_{ij}J_{ij}$ are uniform for any configuration of the global fluxes. Therefore, the difference in the ground-state energy observed for finite systems entirely results from the discrepancy of the ${\bs k}$-grid in the first Brillouin zone when periodic or antiperiodic boundary conditions are used to compute $g_{ij}$ and $J_{ij}$. For gapped systems it is expected that the difference depends exponentially on the circumference \cite{Senthil:2001,Ioselevich:2002} which is consistent with our numerical results, see Fig.~\ref{fig:PD_TO}(b). Besides the four-fold topological degeneracy on the torus, the emergent $Z_2$ gauge structure should also imply a topological term $S_{\rm topo}=-{\rm log}\, 2$ in the entanglement entropy \cite{Yao:2010, Grover:2011}.

\begin{figure}
\centering
\includegraphics[width=1\linewidth]{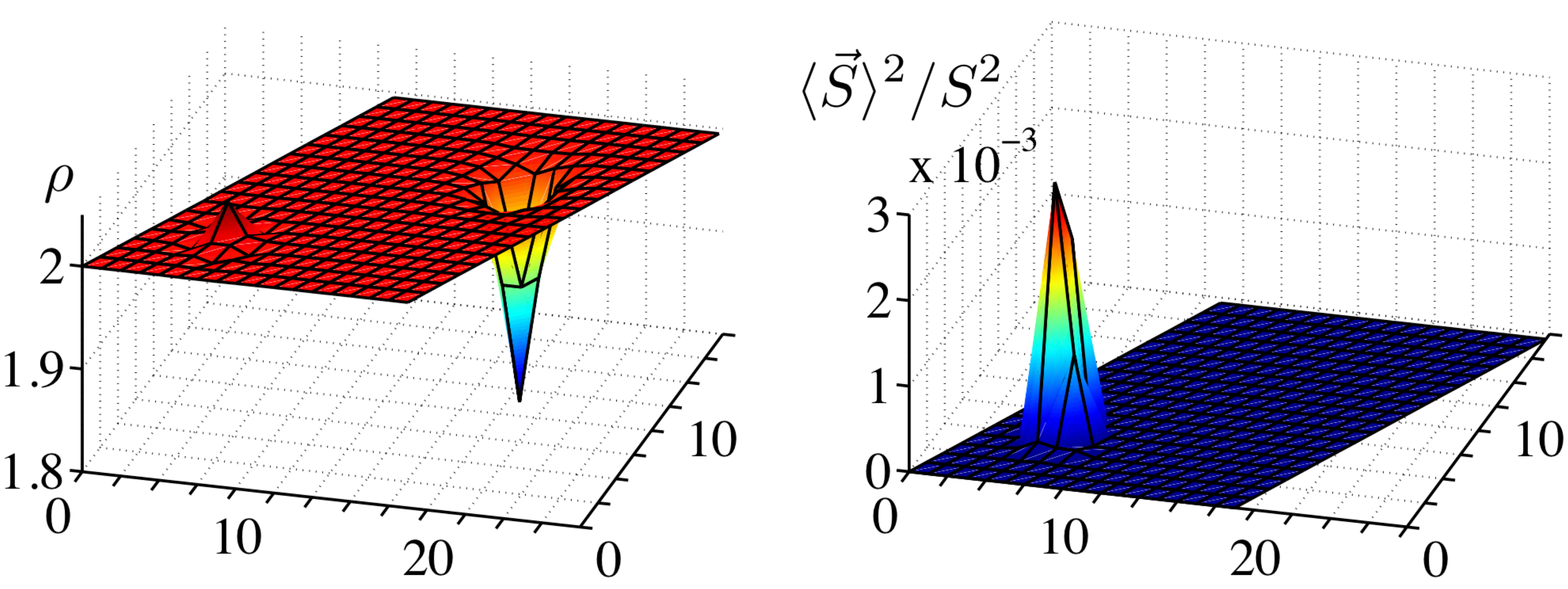}
\caption{(Color online.) Spin-charge separation in the QSH$^*$. Shown is the self-consistent local electron density $\rho$ and the square of the local spin density, $\langle\vec{S}\rangle^2/S^2$, averaged over the Wigner-Seitz cell of the honeycomb lattice. The parameters are $U=21t$, $t_1=0.2t$ and $t_2=0.8t$, and a single hole has ben doped into the half-filled system.}
\label{fig:fractionalization}
\end{figure}

In addition, a special class of excited states exist in the QSH$^*$ phase. They are characterized by the presence of localized $Z_2$ fluxes, $\prod_{\hexagon}{\rm sign}(g_{ij})=-1$, for two hexagons connected by a string of flipped bonds, see Fig.~\ref{fig:fluxpair}. We have validated that the string is not observable in physical (gauge invariant) quantities and have used an approach similar to Ref.~\cite{Ruegg:2011} to confirm that the $Z_2$ fluxes can be separated without cost in energy. This is consistent with the observation of a topological degeneracy and clears the way for the electron fractionalization: because of the non-trivial topological band properties of $H_f$, a $Z_2$ flux introduces a single Kramer's pair into the band gap \cite{Qi:2008,Ran:2008}. There are four different states associated with an isolated $Z_2$ flux: the chargeons (doublon and holon) with charge $\pm e$ but no spin and the charge-neutral Kramer's doublet formed by the spinons. Since two $Z_2$ fluxes can be separated without cost in energy, {\it the chargeons and spinons are elementary excitations of the QSH$^*$ insulator}. Typically, we find that their static energy is comparable to a particle-hole excitation but the relative order depends on details. Because the spin-rotation is fully broken, the spin (or any of its projections) is not a well-defined quantum number. However, following Ref.~\cite{Qi:2008}, the fermion parity operator $\mathcal{I}=(-1)^{\sum_in_i}$ and the time reversal $\mathcal{T}$ still allows to define the generalized spinon $|\psi_s\rangle$ and chargeon $|\psi_c\rangle$ states: $\mathcal{I}|\psi_s\rangle=|\psi_s\rangle$ and $\mathcal{T}^2|\psi_s\rangle=-|\psi_s\rangle$ while $\mathcal{I}|\psi_c\rangle=-|\psi_c\rangle$ and $\mathcal{T}^2|\psi_c\rangle=|\psi_c\rangle$. Regardless of this subtlety, we numerically find that the fractionalization of the electron is still visible when measuring the local electron and spin density, as shown in Fig.~\ref{fig:fractionalization} for a spinon-holon pair.

{\it Gauge fluctuations --} 
Let us now {\it qualitatively} discuss the effect of dynamical fluctuations around the uniform mean-field solution in the QSH$^*$ phase. Because the fluctuations in the {\it magnitude} of the real mean-field values $g_{ij}$ or $J_{ij}$ are gapped, they are not expected to qualitatively change the low-energy behavior and we neglect them in the following. Instead, we focus on the gauge fluctuations. These are fluctuations of the {\it sign} of $g_{ij}$ and $J_{ij}$ which restore the gauge symmetry. The constant amplitude approximation \cite{Senthil:2000} adapted here is expected to be good in the low-energy limit deep inside the QSH$^*$ phase where the length scale associated with a $Z_2$ flux configuration is comparable to the lattice spacing. In the simplest choice consistent with the $Z_2$ gauge structure we allow the mean fields to fluctuate in the following way: for $(i,j)$ nearest neighbors, $\tilde{g}_{ij}=g\tau_{ij}^z$ and $\tilde{J}_{ij}=J\tau_{ij}^z$; for $(i,j)$ second neighbors, $\tilde{g}_{ij}=g'\tau_{ik}^z\tau_{kj}^z$ and $\tilde{J}_{ij}=J'\tau_{ik}^z\tau_{kj}^z$, where $k$ is nearest neighbor to both $i$ and $j$ and the Ising variables $\tau_{ij}^z=\pm 1$ live on the nearest-neighbor bonds. The above relation between the sign of the first- and second-neighbor couplings is also found numerically for inhomogeneous static solutions, see Fig.~\ref{fig:fluxpair}(a). With this particular set of fluctuations, the resulting theory describes pseudoparticles and slave spins minimally coupled to a $Z_2$ gauge field. The physical subspace lies in the gauge invariant sector of the theory defined by
\begin{equation}
\tilde{u}_i=1,\quad{\rm where}\quad \tilde{u}_i=\left(-1\right)^{s_i^z-S+n_i}\prod_{j(i)}\tau_{ij}^x,
\label{eq:Gauss}
\end{equation}
and $j(i)$ labels the nearest-neighbor sites of $i$. Indeed, the local operators $\tilde{u}_i$ generate the gauge transformations: $\tilde{u}_if_{i\alpha}^{(\dag)}\tilde{u}_i=-f_{i\alpha}^{(\dag)}$, $\tilde{u}_is_{i}^{x}\tilde{u}_i=-s_{i}^{x}$ and $\tilde{u}_i\tau_{ij}^{z}\tilde{u}_i=-\tau_{ij}^{z}$. In order to obtain a physical picture of the ``Gauss law" Eq.~\eqref{eq:Gauss} we denote a site with $u_i=(-1)^{n_i+s_i^z-S}=-1$ as being occupied by a $Z_2$ charge (both pseudo fermions and slave spins carry a $Z_2$ charge). Similarly, we can define a $Z_2$ electric field which is finite on bonds with $\tau_{ij}^x=-1$. Equation~\eqref{eq:Gauss} now implies that $Z_2$ charges are created in pairs and are connected by the $Z_2$ electric field if created on different sites. The  physical subspace of the original $Z_2$ representation, i.e.~$u_i=1$, is identified as the $Z_2$ charge free subspace in the gauge invariant sector of the effective gauge theory.

\begin{figure}
\centering
\includegraphics[width=0.9\linewidth]{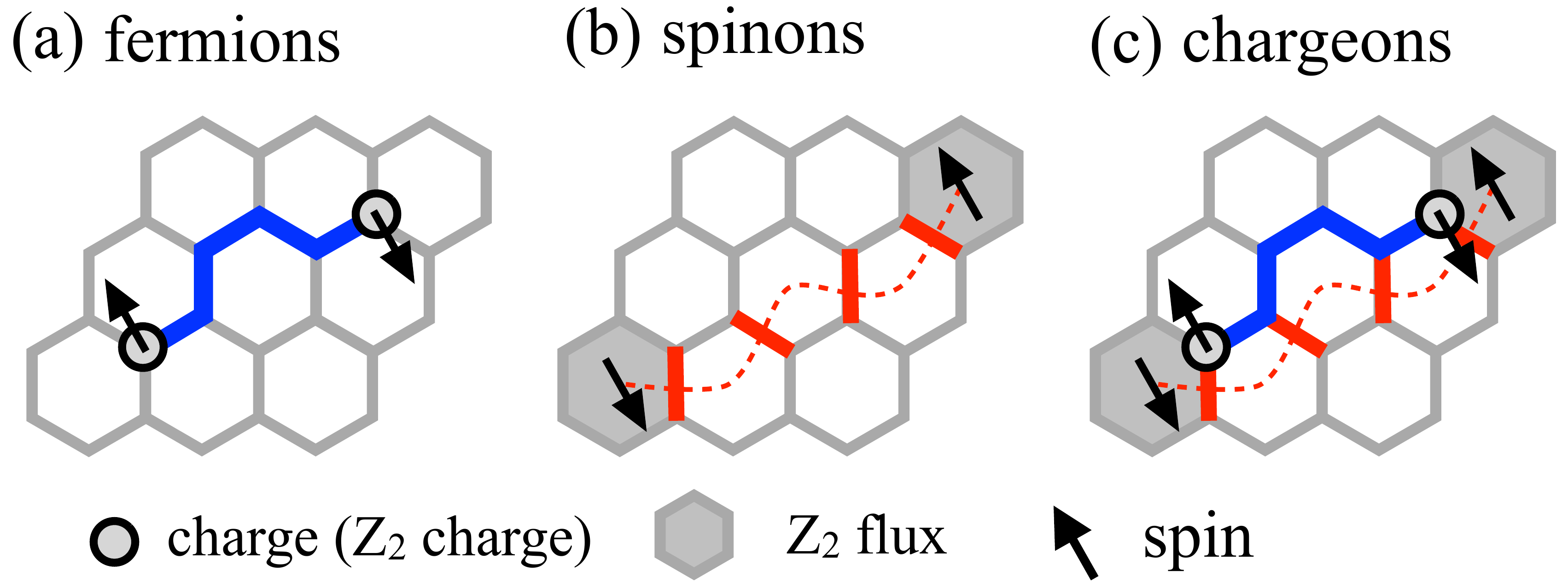}
\caption{(Color online.) Excitations in the QSH$^*$. (a) The fermionic pseudoparticles carry an electronic charge and a spin as well as a $Z_2$ charge. (b) Spinon exciations carry a (generalized) spin but no charge and are bosons. (c) Doublons and holons have charge but no spin and are also bosons. The mutual braiding statistics is semionic.}
\label{fig:excitations}
\end{figure}

To proceed, we note that the slave-spin excitations are gapped in the QSH$^*$ phase. At least conceptually, it is therefore possible to ``integrate them out" \footnote{A path integral representation of the current theory similar to Ref.~\cite{Senthil:2000} is discussed in a forthcoming article.}. The resulting theory consists of the fermionic pseudoparticles which are coupled to a dynamical $Z_2$ gauge field and the physical subspace is now given by $(-1)^{n_i-1}=\prod_{j(i)}\tau_{ij}^x$ \cite{Ran:2008}. In the process of integrating out the slave spins, new terms are generated. These terms govern the dynamics of the gauge field and have to be consistent with the gauge symmetry. In lowest order the resulting Hamiltonian is therefore given by
\begin{equation}
H_{\tau}=-K_{\tau}\sum_{\hexagon}\prod_{\hexagon}\tau_{ij}^z-I\sum_{\langle i,j\rangle}\tau_{ij}^x-G\sum_i\prod_{j(i)}\tau_{ij}^x.
\label{eq:Htau}
\end{equation}
From the analysis of the static mean-field solutions, we expect that the parameters $K_{\tau}$, $I$, and $G$, are non-negative. There are two different phases of $H_{\tau}$, a confining phase if $I$ dominates and a deconfining phase if $K_{\tau},G$ dominates over $I$ \cite{Wen:2003}. In the QSH$^*$ phase, the gauge theory is in the deconfining phase.
As shown by Wen \cite{Wen:2003}, an instructive picture applies in this case: the ground state of Eq.~\eqref{eq:Htau} can be viewed as a string-net condensation of closed strings. Excitations are open strings with emergent quasiparticles at their ends. There are three different types of excitations which involve either a $Z_2$ charge (signaled by $\prod_{j(i)}\tau_{ij}^x=-1$), a $Z_2$ flux (signaled by $\prod_{\hexagon}\tau_{ij}^z=-1$) or both. $Z_2$ charge and $Z_2$ flux are both bosonic excitations while the bound state formed of both is a fermion. Moreover, when a $Z_2$ charge encircles a $Z_2$ flux, a phase factor of $\pi$ is picked up. We now use these results for the pure gauge theory to obtain the exchange and mutual braiding statistics of the low-lying excitations in the QSH$^*$ phase. For this purpose, the coupling of the gauge field to the matter field of the pseudoparticles has to be considered. Specifically, this means that we attach a fermionic pseudoparticle to every $Z_2$ charge and a (generalized) spin to every $Z_2$ flux. Hence, the elementary excitations are (a) the fermionic pseudoparticles with spin, charge and $Z_2$ charge, (b) the bosonic spinons with $Z_2$ flux and generalized spin and (c) the bosonic chargeons with charge, $Z_2$ charge, $Z_2$ flux but no spin, see Fig.~\ref{fig:excitations}. Upon mutual braiding, a phase $\pi$ is picked up and the mutual braiding statistics is semionic. For the spinons and chargeons this result agrees with Ref.~\cite{Ran:2008}. In our discussion we have additionally included the fermionic pseudoparticles. Note that although the emergent fermion carries the degrees of the electron it is distinct from it by the additional $Z_2$ charge.

{\it Physical response of the QSH$^*$ insulator --}
We are now in a position to discuss the experimental signatures of the QSH$^*$ phase which allow one to distinguish this phase from the trivial band insulator and the (interacting) QSH phase. We first address the edge properties. In analogy to the QSH insulator, the mean-field theory predicts that the fermionic pseudoparticles form a single bidirectional pair of gapless edge states related by time reversal symmetry. The slave spins, on the other hand, are gapped everywhere. In an edge-state theory for the pseudoparticles, we find that sufficiently strong residual interactions have the potential to open a gap by spontaneously breaking the time-reversal symmetry at zero temperatures, again in analogy to the interacting QSH insulator \cite{Wu:2006,Xu:2006}. In the following, we will focus on the more interesting case where the time-reversal symmetry is preserved and the gapless edge modes survive (see Refs.~\cite{Hohenadler:2011,Imada:2011, Lee:2011} in the context of the interacting QSH phase). Because the pseudoparticle is {\it not} proportional to the electron in the QSH$^*$ phase ($Z=\langle s_i^x\rangle/S=0$), we expect that the edge spectrum looks gapped in a single-particle tunneling experiment. However, the gapless character should manifest itself in the power-law decay of the physical charge and spin correlation functions (in the mean-field theory, $\langle c_{i\alpha}^{\dag}c_{i\beta}c_{j\gamma}^{\dag}c_{j\delta}\rangle\approx\langle f_{i\alpha}^{\dag}f_{i\beta}f_{j\gamma}^{\dag}f_{j\delta}\rangle$). In principle, the charge correlation function can be (indirectly) measured in a Coulomb drag experiment \cite{Zyuzin:2010} and the spin correlation function with neutron scattering. 

The presence of gapless edge states protected by time-reversal symmetry distinguishes the QSH$^*$ phase from a trivial insulator. But interestingly, the QSH$^*$ phase is a trivial insulator with respect to its single-particle response in the sense of Ref.~\cite{Qi:2008}. More precisely, let us study the response to an {\it external} $\pi$-flux. While in the QSH insulator an isolated $\pi$-flux tube binds a Kramer's doublet leading to spin-charge separation \cite{Qi:2008,Ran:2008}, the emergent $Z_2$ gauge field in the QSH$^*$ phase can completely screen the external $\pi$-flux. Indeed, within the static mean-field description, we numerically confirmed that the ground state in the presence of an external $\pi$-flux does not have isolated mid-gap states. Consequently, an adiabatic insertion of a $\pi$-flux does not induce spin-charge separation. In particular, we conclude that the topological $Z_2$-invariant defined by the parity of the charge pumped towards the isolated flux tube during the adiabatic insertion of a {\it spin}-$\pi$-flux \cite{Qi:2008} is trivial in the QSH$^*$ phase (as opposed to QSH insulator).

{\it Conclusions --}
In conclusion, we have provided a framework to discuss the exotic QSH$^*$ phase within a self-consistent theory. We have identified a parameter regime in a strongly interacting physical model where this phase is stabilized and accessible to other (numerical) methods. As in the conventional QSH insulator, the ground state of the QSH$^*$ phase does not break any symmetry, is gapped in the bulk but has gapless edge modes protected by time-reversal symmetry. However, it does not show the quantum spin Hall effect. The picture of string-net condensation allowed us to derive the braiding statistics of the emergent fermionic quasiparticles and the spin-charge separated $Z_2$-flux excitations.
\acknowledgments
We would like to thank T. Senthil for an early discussion on the $Z_2$ representation and M. Kargarian and J. Wen. We acknowledge financial support through ARO grant W911NF-09-1-0527 and NSF grant DMR-0955778.

%

\end{document}